\documentclass[11pt]{article}
\usepackage{latexsym,amsmath,amssymb,xfrac}
\usepackage{epstopdf,xcolor}
\usepackage{hyperref}
\newcommand{\pa}{{\partial}}

\makeatletter
\renewcommand\section{\@startsection{section}{1}{\z@}
  {-3.5ex \@plus -1ex \@minus -.2ex}%
    {2.3ex \@plus.2ex}
    {\vskip5pt\normalfont\large\bfseries}}
\renewcommand\subsection{\@startsection{subsection}{2}{\z@}%
    {-3.5ex \@plus -1ex \@minus -.2ex}%
    {1.3ex \@plus.2ex}%
    {\noindent\normalfont\normalsize\underline}}%
\makeatother

\numberwithin{equation}{section}

\begin{document}
\thispagestyle{empty}

\begin{center}
  \noindent {\bf\LARGE Computing the energy of some timelike singularities}\\[15pt]
\noindent
Fernando~Ruiz~Ruiz\footnote{\href{mailto:ferruiz@ucm.es}{ferruiz@ucm.es},
  \href{https://orcid.org/0000-0003-1571-2468}{ORCID: 0000-0003-1571-2468}}\\[2pt]
     \noindent {\sl Departamento de F\'{\i}sica Te\'orica, Facultad de
       Ciencias F\'{\i}sicas, \\Universidad Complutense de Madrid,
       28040 Madrid Spain}
\end{center}

\vskip 30pt
{\leftskip=20pt\rightskip=20pt 

  \noindent

  The gravitational energies of timelike singularities in some
  asymptotically AdS spaces and in the Schwarzschild black hole with
  negative mass are explicitly computed. In all cases considered, the
  energy is finite and positive, depends on the geometry around the
  singularity, and compensates negative energy contributions from the
  asymptotic boundary, rendering the total gravitational energy
  positive. The definition used for energy is the on-shell Hamiltonian
  of the physical classical action proposed by Hawking and
  Horowitz~\cite{Hawking-Horowitz}. To deal with the singularity,
  spacetime is cut off by a timelike regulating boundary surrounding
  the singularity that is compatible with a foliation of spacetime in
  constant-time slices. It is shown that the total energy of static
  AdS-Kasner solutions with timelike Kasner singularities is equal to
  the energy of the planar AdS black hole, and that the Schwarzschild
  black hole with negative mass has the same
  energy as Minkowski space.\\[5pt]
  
  \noindent {\sc keywords:} timelike singularity, energy, AdS-Kasner 
  metric, negative/positive mass black hole.

\par}

\vskip 30pt

\section{Introduction}

Understanding spacetime singularities and the physics near them has
been an ongoing issue in general relativity for the last fifty
years. On the one hand, it is widely believed that they must be
resolved and that new ultraviolet physics must be formulated for
that. On the other, it is acknowledged that extensions of general
relativity including higher curvature interactions will have solutions
with null singularities~\cite{HM-value-singularities}. Timelike
singularities also play an important part, for they are necessary to
eliminate unphysical negative energy
solutions~\cite{HM-value-singularities}. Consider for example a black
hole with `negative mass'. If the timelike singularity at its center
were smoothed out without modifying its asymptotic behaviour, one
would end up with a spacetime with arbitrarily negative energy. This
observation has implications for quantization, for a classical theory
with such solutions is unlikely to head to a quantum theory with a
stable ground state~\cite{HM-value-singularities}. Here we adopt a
somewhat different but complementary point of view and propose to
regard the timelike singularity as providing positive energy that when
added to the negative asymptotic energy gives a positive total
gravitational energy.

Our approach is based on the observation that, since timelike
singularities are localized in a spatial region that can be avoided at
all times, they can be considered as static, so it might be possible
to define an energy for them. The energy would of course depend on the
the geometry of the solution near the singularity. In this note we
realize this idea by computing the energy of some asymptotically AdS
and asymptotically flat spaces with time-translational symmetry and a
timelike singularity.

To deal with the singularity, we cut off spacetime with a timelike
regulating boundary $\Sigma^{\epsilon}$ that surrounds the singularity
and regard it as part of the spatial boundary on the same footing as
the asymptotic boundary $\Sigma^\infty$.  We combine this
regularization with the defintion by Hawking and
Horowitz~\cite{Hawking-Horowitz} of energy for an asymptotically AdS
or asymptotically flat space with time-translational invariance as the
the on-shell Hamiltonian of the physical classical action.  For this
to be possible, the regulated spacetime must admit a foliation in
constant-time slices $\{\Sigma_t\}$, which is the case for all
instances considered here. Every slice $\Sigma_t$ then has boundaries
at $\Sigma^{\infty}$ and $\Sigma^{\epsilon}$, and the boundary
$\partial M$ of the regulated spacetime is formed by initial and final
constant-time slices $\Sigma_{t_1}$ and $\Sigma_{t_2}$, the asymptotic
boundary $\Sigma^{\infty}$ and the boundary $\Sigma^{\epsilon}$ near
the singularity.  The same arguments as those used by Hawking and
Horowitz~\cite{Hawking-Horowitz} give the regulated energy as the sum
$E=E_{\infty}+E_{\textnormal{\sc sing}}$ of a contribution $E_\infty$
from the asymptotic boundary and a contribution
$E_{\textnormal{\sc sing}}$ from the boundary surrounding the
singularity. Here we compute $E_{\textnormal{\sc sing}}$ for fixed
regulator $\epsilon$ and then set $\epsilon$ to zero.

We test this approach with two examples. In Section 2, we consider a
family of asymptotically AdS metrics~\cite{Ren} that includes as
particular cases the AdS planar black hole, the AdS soliton and AdS spaces
with Kasner timelike singularities. Our results show that there are
infinite solutions among them whose asymptotic contribution $E_\infty$
to the energy is arbitrarily negative, but whose contribution
$E_{\textnormal{\sc sing}}$ from the singularity is always positive,
well-defined
and such that the total energy $E_\infty+E_{\textnormal{\sc sing}}$ is
positive. The only two geometries in the family that do not have a
timelike singularity, so their energy is given by $E_\infty$, are the
AdS soliton and the AdS planar black hole, for which the energies in
the literature are recovered~\cite{HM-soliton}. Our computations show
that the solutions with timelike singularity have the same energy as
the planar AdS black hole. Section 3 is much shorter and discusses the
Schwarzschild metric with arbitrary mass parameter. As is well known,
the black hole with positive mass has positive energy relative to
Minkowski space.  For the black hole with negative mass, it will be
shown that the negative energy from the asymtotic region is canceled
by the enery from the timelike singularity at its center, thus giving
zero total energy relative to Minkowski space.

\section{Energy from timelike singularities in AdS-Kasner spaces}

The Einstein equations with negative cosmological constant
\begin{equation}
  R_{\mu\nu} -\frac{1}{2}\,g_{\mu\nu}\,R+g_{\mu\nu}\Lambda = 0
 \label{EE}
\end{equation}
have the following family of static radial solutions in $n\!+\!1$
dimensions~\cite{Ren},
\begin{equation}
  ds_{n+1}^2 = \frac{\ell^2}{z^2}\,\bigg[\! 
  - f({z})^{p_0\!}\,  d{t}^2
  + \frac{d{z}^2}{f(z)}  
  + \sum_{i=1}^{n-1} f({z})^{p_i\!}\,(dx^i)^2\,\bigg]\,.
\label{metric-z}
\end{equation}
Here $\ell$ is the AdS radius, related to the cosmological constant
through $\Lambda=-{n\,(n-1)}/{2\ell^2}$, $t$ and $z>0$ are time
and radial coordinates, $x^i$ are transverse coordinates with
periodicities $L_i$, the function $f(z)$ is given by
\begin{equation}
  f({z})=1-\bar{z}^{n},\qquad \bar{z}=\frac{z}{z_0}\,,
  \label{fz}
\end{equation}
with $z_0$ an integration constant, and the
parameters $p_a$ ($a=0,1,\ldots n-1$) satisfy the Kasner-type
conditions
\begin{equation}
  \sum_{a=0}^{n-1} p_a = \sum_{a=0}^{n-1}p^2_a  =1\,.
  \label{conditions}
\end{equation}
All other integration constants can and have been absorbed in the
periodicities $L_i$, so the $L_i$ themselves are integration
constants.\footnote{The solutions in eq.~(\ref{metric-z}) can be
  obtained as follows. Write the metric ansatz in the Fefferman-Graham
  gauge
\begin{equation}
  ds_{n+1}^2 = \frac{\ell^2}{w^2}\,\big[\!- e^{2A_0(w)} dt^2 + dw^2 
    + \sum_{i=1}^{n-1} e^{2A_i(w)} (dx^i)^2 \big]\,.
  \label{metric-ansatz}
\end{equation}
In these coordinates the Einstein equations
are very easily solved, the solution being
\begin{equation}
  ds_{n+1}^2 = \frac{\ell^2}{w^2}\, \bigg\{ dw^2 
  + {\left( 1+\bar{w}^{n} \right)}^{4/n} \bigg[
  - {\left( \frac{1-\bar{w}^{n}}{1+\bar{w}^{n}} \right)}^{2p_0}  dt^2 
  + \sum_{i=1}^{n-1}\, k_i^2\, {\left( \frac{1-\bar{w}^{n}}
      {1+\bar{w}^{n}} \right)}^{2p_i} \,{(dx^i)}^2 \bigg] \bigg\},
\label{metric-w}
\end{equation}
with $\bar{w}=w/z_0$, $p_a$ as in eqs.~(\ref{conditions}) and $k_i$
arbitrary integration constants. To go from this expression for the
metric to that in eq.~(\ref{metric-z}), introduce a radial coordinate
$z$ through the differential equation
\begin{equation}
  \frac{dw^2}{w^2}
       = \frac{dz^2}{z^2\,f({z})}\,,
  \label{radial}
\end{equation}
substitute its solution $z^n=4w^n/(1+\bar{w}^n)^2$ in
eq.~(\ref{metric-w}) and rescale the coordinates $x^i$.} For
convenience we assume that the transverse coordinates $x^i$ are
compact. The noncompact case can be retrieved from the results below.

At $z\to 0$ the metrics (\ref{metric-z}) are asymptotically AdS. To study their
behaviour near $z=z_0$, make the change
$ \bar{z} = 1-({n\rho^2}/{4\ell^2})$ and expand in powers of
$\rho/\ell$. This gives
\begin{equation}
  ds^2_{n+1} (\rho\to 0)= d\rho^2 + \frac{\ell^2}{z_0^2}\, \bigg[\! 
  - \bigg(\frac{n\rho}{2\ell}\bigg)^{2p_0} dt^2
  + \sum_{i=1}^{n-1} \bigg(\frac{n\rho}{2\ell}\bigg)^{2p_i} (dx^ i)^2 \,\bigg]\,.
    \label{Kasner}
\end{equation}
For generic values of $p_a$, the metric (\ref{Kasner}) describes a
Kasner spacetime~\cite{Kasner}, which is known to solve the vacuum
Einstein equations and to have a naked timelike singularity at
$\rho=0$. Along directions $x^a$ with $p_a>0$ space looks like a
crunch, whereas along those with $p_a<0$ it resembles a rip.  We term
the corresponding solutions~(\ref{metric-z}) AdS Kasner. They have
been regarded~\cite{Ren}\,\cite{Shaghoulian} as timelike variants of
the spacelike singularities of Belinski, Khalatnikov and
Lifshitz~\cite{BKL}, and their feasibility as flows from pure AdS
spacetime at $z\to 0$ to Kasner spaces at $z=z_0$ has been studied by
several authors~\cite{Ren}\,\cite{Shaghoulian}\,\cite{Engelhardt}.
There are two choices for the parameters $p_a$ for which the
metric~(\ref{metric-z}) is not singular at $z=z_0$. They correspond to
the planar AdS black hole and the AdS soliton.

\medskip \underline {The planar AdS black-hole}. Take $p_0\!=1$, hence
$p_1\!=\ldots=p_{n-1}\!=0$.  The solution~(\ref{metric-z}) describes
then an $(n\!+\!1)$--dimensional planar AdS black hole with infinity
at $z\to 0$, an event horizon at $z=z_0$, and a cloaked spacelike
singularity at $z\to \infty$. The black hole mass is given in terms of
$z_0$ by
$M_{\textnormal{\sc bh}}= (n-1)\,V_{n-1}\ell^{n-1}/16\pi G z_0^n$,
where $V_{n-1}=\Pi_i L_i$ is the coordinate volume of the surface
parameterized by $x^i$. In this case, the metric in eq.~(\ref{Kasner})
reduces to the product of a two-dimensional Rindler metric with a flat
$(n\!-\!1)$--dimensional metric,
\begin{equation}
  ds^2=-\bigg(\frac{n\rho}{2z_0}\bigg)^2 dt^2 + d\rho^2
  + \frac{\ell^2}{z_0^2}\,\sum_{i=1}^{n-1}{(dx^i)}^2,
  \label{Rindler}
\end{equation}
which is the metric of a planar black hole in the outer neighborhood
of its horizon.

\medskip \underline{The AdS soliton.} Consider now that it is one of
the `spatial' parameters $p_i$ that takes the value~1; say
$p_1\!=1$. Thence $p_0\!=p_2\!=\cdots=p_{n-1}\!=0$. The
$(\rho,x^1)$-sections of the metric~(\ref{Kasner}) have now an
apparent conical singularity at $\rho=0$. This is
eliminated~\cite{HM-soliton}\,\cite{HWY} by taking the coordinate
$x^1$ to be compact with periodicity $L_1$ given by
\begin{equation}
  L_1=\frac{4\pi z_0}{n}\,.
    \label{constants-sol}
\end{equation}
In terms of an angle coordinate $\chi=2\pi x^1/L_1$ taking values in
$[0,2\pi]$, the metric (\ref{Kasner}) near $\rho=0$ is regular and
takes the form
 \begin{equation}
   ds^2= d\rho^2 + \rho^2 d\chi^2 + \frac{\ell ^2}{z_0^2} \Big[\, -dt^2
   + \sum_{i=2}^{n-1}{(dx^i)}^2\,\Big]\,,
  \label{regular}
\end{equation}
The full metric~(\ref{metric-z}) reproduces in this case the
Horowitz-Myers AdS soliton~\cite{HM-soliton},
\begin{equation}
  ds^2_{\textnormal{\sc sol}}=\frac{\ell^2}{\bar{z}^2}\,
  \bigg[ -d\bar{t}^2 + \frac{d\bar{z}^2}{f(z)} +
  \frac{4}{n^2}\,f(z)\,d\chi^2 +
  \sum_{i=2}^{n-1}{(d\bar{x}^i)}^2\,\bigg]\,,
   \label{soliton}
\end{equation}
where dimensionless coordinates $\bar{t}=t/z_0,\,\bar{z}$ and
$\bar{x}^i=x^i/z_0$ have been introduced.

The integration constant $z_0$ specifies the location of the timelike
singularity in the AdS-Kasner solutions, whereas in the black hole it
determines the black hole mass, but otherwise it is arbitrary. In
other words, in these two cases, the integration constants $L_i$ and
$z_0$ are independent, so fixing the conformal boundary does not fix
$z_0$. For the soliton, however, $z_0$ is given in terms of $L_1$ by a
regularity requirement, so fixing the conformal boundary fixes $z_0$.

We want to calculate the energies of the solutions given in
eq.~(\ref{metric-z}). To do this, we first review the Hawking-Horowitz
expression~\cite{Hawking-Horowitz} for the energy of an asymptotically
AdS/flat spacetime and then modify it so as to take care of the
singularity at $z=z_0$. Consider an asymptotically AdS or
asymptotically flat solution $g_{\mu\nu}$ to the Einstein equations
and denote by $\tilde{g}_{\mu\nu}$ the static reference AdS or flat
background to which it asymptotically approaches. Call $\Sigma^\infty$
to the boundary near infinity at which they agree. Assume that
spacetime $M$ is foliated in spacelike surfaces $\{\Sigma_t\}$ by a
timelike Killing vector field of norm squared $-N^2$, and that the
only boundary of the surfaces $\{\Sigma_t\}$ is at
$\Sigma^\infty$. The spacetime boundary $\partial M$ is formed by
initial and final time surfaces $\Sigma_{t_1}$ and $\Sigma_{t_2}$, and
by $\Sigma^{\infty}$.  The foliation $\{\Sigma_t\}$ induces a
foliation of $\Sigma^{\infty}$ in the surfaces $\{S^{\infty}_t\}$ that
result from the intersection of $\Sigma^{\infty}$
with~$\{\Sigma_t\}$. The energy of the solution $g_{\mu\nu}$ relative
to the background $\tilde{g}_{\mu\nu}$ is defined as the on-shell
Hamiltonian for the physical classical action\footnote{The physical
  classical action is the difference
  $S_{\textnormal{\sc phys}}=S[g]-S[\tilde{g}]$, where $S[g]$ is the
  sum of the Einstein-Hilbert and Gibbons-Hawking-York~\cite{GHY}
  terms.  $S_{\textnormal{\sc phys}}$ is finite for the class of
  solutions $g_{\mu\nu}$ that asymptotically approach
  $\tilde{g}_{\mu\nu}$.} and is given by
\begin{equation}
  E =-\,\frac{1}{8\pi G} \bigg(
  \int_{S^\infty_t} \!\!d^{n-1\!}x\, \sqrt{|\sigma|}N{\kappa} -
  \int_{\tilde{S}^\infty_t} \!\!d^{n-1\!}\tilde{x}\,
  \sqrt{|\tilde\sigma|}{N}\tilde{\kappa}\bigg) ,
  \label{HH-energy}
\end{equation}
where $\sigma_{ij}$ is the metric of ${S}^{\infty}_t$ and $\kappa$ is
the trace of its intrinsic curvature in $\Sigma_t$. The reference
background tilded quantities are defined likewise:
$\tilde{\sigma}_{ij}$ is the metric on $\tilde{S}^\infty_t$ and
$\tilde{\kappa}$ is the trace of the extrinsic curvature of
$\tilde{S}^\infty_t$ in $\tilde{\Sigma}^\infty_t$. By construction
${S}^\infty_t$ and $\tilde{S}^\infty_t$ have the same intrinsic
geometry, and the lapse function $N$ is the same in both integrals, so
the AdS-Kasner and the reference background energies are compared in
the same way (see ref.~\cite{Hawking-Horowitz} for details).

In ref.~\cite{Hawking-Horowitz} it is proved that
this expression for the energy agrees with that of Abbott and
Deser~\cite{Abbott-Deser} for asymptotically AdS spaces, and with the
ADM energy~\cite{ADM} for asymptotically flat spaces. It is also shown
there that it is valid for (maximal extensions of) black holes with
one event horizon but two asymptotic regions. All this implies that
eq.~(\ref{HH-energy}) can be readily used to determine the energies of
the AdS planar black hole and the AdS soliton above. These were
computed in ref.~\cite{HM-soliton}, with the result that, whereas the
black hole has positive energy, the soliton has negative energy, which
led to a new positive energy conjecture.

For the AdS-Kasner solutions in eq.~(\ref{metric-z}), the reference
background is AdS space and the spatial boundary $\Sigma^\infty$ is at
$z=0$. To deal with their timelike singularity at $z=z_0$, we
cut off spacetime along a surface $\Sigma^z$ of constant-$z$ and take
$z=\epsilon$, with $0\!\ll\! {\epsilon}\!<\!z_0$, metric given by
\begin{equation}
  ds^2_n(\Sigma^z) = \frac{\ell^2}{z^2}\,\bigg[ \, -f(z)^{p_0} dt^2
     + \sum_{i=1}^{n-1} f(z)^{p_i} (dx^i)^2\,\bigg] \,.
\end{equation}
Now we regard $\Sigma^\epsilon$ as regulating surface\footnote{The
  volume of $\Sigma^\epsilon$ is
  $\ell^n f(\epsilon)^{1/2}/\epsilon^n$, which is nonzero for
  $\epsilon <z_0$. As $\epsilon\to z_0^-$, the volume approaches zero
  and the singularity is incorporated into spacetime.}  that is part
of the boundary on the same footing as the asymptotic boundary
$\Sigma^\infty$. The vector field $\pa_t$ is timelike and Killing,
has norm squared
\begin{equation}
  -N^2=-\frac{\ell^2}{z^2}\,{f(z)}^{p_0} ,
  \label{Killing}
\end{equation}
and foliates the regulated spacetime in spacelike surfaces
$\{\Sigma_t\}$ of constant $t$ with metric
\begin{equation}
  ds^2_n(\Sigma^t) = \frac{\ell^2}{z^2}\,\bigg[ \, \frac{dz^2}{f(z)} 
     + \sum_{i=1}^{n-1} f(z)^{p_i} (dx^i)^2\,\bigg] \,.
\end{equation}
For fixed $\epsilon <z_0$, we have a well-defined problem: the
boundaries $\Sigma^{\epsilon}$ and $\Sigma^\infty$ are both timelike
and are foliated in surfaces $\{S^{\epsilon}_t\}$ and $\{S^\infty_t\}$
that result from the intersections of $\Sigma^{\epsilon}$ and
$\Sigma^\infty$ with $\{\Sigma_t\}$. The slices $\Sigma_t$ do not
intersect and meet both $\Sigma^{\infty}$ and $\Sigma^{\epsilon}$
orthogonally. The assumption in ref.~\cite{Hawking-Horowitz} that the
only boundary of the constant-time surfaces $\{\Sigma_t\}$ is
$\Sigma^\infty$ no longer holds, since now every $\Sigma_t$ has a
boundary at $\Sigma^\infty$ and another one at
$\Sigma^{\epsilon}$. The spacetime boundary is formed by
$\Sigma_{t_1},\,\Sigma_{t_2}$, $\Sigma^\infty$ and
$\Sigma^{\epsilon}$. Performing at fixed $\epsilon<z_0$ the same
arguments as in ref.~\cite{Hawking-Horowitz} and sending
$\epsilon\to z_0$ in the on-shell Hamiltonian give for the energy of
the solutions (\ref{metric-z}) relative to the AdS background the sum
\begin{equation}
  E=E_\infty+ E_{\textnormal{\sc sing}}
  \label{energy}
\end{equation}
of two contributions
\begin{equation}
  E_\infty=  \lim_{z\to 0} E(z)~~\textnormal{and}~~ 
  E_{\textnormal{\sc sing}} 
         = \lim_{\epsilon\to z_0^-}  E(\epsilon)\,,
\label{energy-asy-sin}
\end{equation}
the first one of which is from the asymptotic boundary and the second one
from the singularity. These are obtained by sending $z\to 0$ and
$z=\epsilon\to z_0^-$ in
\begin{equation}
  E(z)=  -\frac{1}{8\pi G} \bigg(
  \int_{S^z_t} \!d^{n-1\!}x\, \sqrt{|\sigma|} N\/\kappa -
  \int_{\tilde{S}^z_t} \!d^{n-1\!}\tilde{x}\, \sqrt{|\tilde{\sigma}|}
                      {N} \tilde{\kappa}\bigg).
\label{E(z)}
\end{equation}
As discussed earlier, the black hole and the soliton energies are made
of only the contribution~$E_\infty$.

Let us compute $E_{\infty}$ and $E_{\textnormal{\sc sing}}$. A surface
$S^z_t$ of constant $z$ in a constant-$t$ slice $\Sigma_t$ is the
product of $n\!-\!1$ circles of radii $\ell{f(z)}^{p_i/2}/z$ and
periods $L_i$. Its metric is
\begin{equation}
   ds^2_{n-1}=  \sigma_{ij}dx^i dx^j
   = \frac{\ell^2}{z^2}\,\sum_{i=1}^{n-1} {f(z)}^{p_i}\,(dx^i)^2,
    \label{metric-sigma}
\end{equation}
and the trace of its extrinsic curvature in $\Sigma_t$ is given by
\begin{equation}
   \kappa =\mp\, \frac{z\,{f(z)}^{1/2}}{2\ell}\,\sigma^{ij}\,\pa_z\sigma_{ij}.
\label{K}
\end{equation}
The upper sign must be taken when calculating $E_\infty$, for then the
surface $S^z_t$ approaches $z=0$ and the exterior unit vector normal
to $S^z_t$ is\, $n^t\!=0,~n^z\!=-\!z\sqrt{f}/2l,~n^i\!=0$.  The lower
sign is to be taken for $E_{\textnormal{\sc sing}}$, since when
$z=\epsilon$ approaches $z_0$ the exterior unit vector normal to
$S^\epsilon_t$ has opposite direction,\,
$n^t\!=0,~n^z\!=z\sqrt{f}/2l,~n^i\!=0$.  Some simple algebra gives for
the first integral in eq.~(\ref{E(z)})
\begin{equation}
  \int_{S^z_t}\!\!d^{n-1}x \,N\sqrt{|\sigma|}\kappa= \pm\, V_{n-1}\, 
  \frac{\ell^{n-1}}{2z^n}\, \bigg[2(n-1) + (2-n-np_0)\,\frac{z^n}{z_0^n}\, \bigg]\,,
   \label{N-sig-K}
\end{equation}
with $V_{n-1}=\Pi_iL_i$ the coordinate volume of the surface
parameterized by $x^i$. It has already been mentioned that the
reference background is AdS.  To calculate the second integral in
eq.~(\ref{E(z)}), we must find a surface $\tilde{S}^z_t$ in the
background with the same intrinsic geometry as $S^z_t$. This is the
product of $n\!-\!1$ circles of radii $\ell/z$ and periods
$\tilde{L}_i$. For $\tilde{S}^z_t$ and $S^z_t$ to have the same
intrinsic geometry, proper distances along circles in $\tilde{S}^z_t$
must be the same as along circles in $S^z_t$. This provides the
condition
\begin{equation}
  \frac{\ell}{z}\,\tilde{L}_i
  = \frac{\ell}{z}\,f(z)^{p_i/2}\,L_i.
\label{constant-i} 
\end{equation}
which gives $\tilde{L}_i$ in terms of $L_i$ for every given $z$. With
this, we have
\begin{align}
  \int_{\tilde{S}^z_t}\!\!d\tilde{x}^{n-1} \,{N}
      \sqrt{|\tilde{\sigma}|}\tilde{\kappa}
  & =   \pm \,\big(\Pi_i\tilde{L}_i\big)\, \frac{\ell^{n-1}}{z^n}\,f(z)^{p_0/2}\,(n-1)
    \nonumber\\[3pt]
  & = \pm \,V_{n-1}\, \frac{\ell^{n-1}}{z^n}\, f(z)^{1/2}\,(n-1)\,.
   \label{N-sig-K-tilde}
\end{align}

Inserting eqs.~(\ref{N-sig-K}) and~(\ref{N-sig-K-tilde}) with upper
signs in eq.~(\ref{E(z)}), and sending in the result $z\to 0$, it is
straightforward to obtain
\begin{equation}
  E_{\infty} =   \frac{\ell^{n-1}V_{n-1}}{16\pi Gz_0^n}\, (np_0-1)\,.
  \label{E-inf}
\end{equation}
To find $E_{\textnormal{\sc sing}}$, insert eqs.~(\ref{N-sig-K})
and~(\ref{N-sig-K-tilde}) with lower signs in eq.~(\ref{E(z)}) and let
$z=\epsilon\to z_0^-$. This yields\
\begin{equation}
   E_{\textnormal{\sc sing}}= \frac{ \ell^{n-1}V_{n-1}}{16\pi Gz_0^n}\,n(1-p_0)\,.
  \label{E-sin}
\end{equation}
Note that the operations leading to $E_{\textnormal{\sc sing}}$ do not
make sense without the regulating boundary $\Sigma^{\epsilon}$
($\epsilon\!<\!z_0$). For $\epsilon=z_0$, spacetime cannot be foliated
in non-intersecting slices $\{\Sigma_t\}$, the surfaces $S^\epsilon_t$
cannot be defined, and the lapse function $N=f(\epsilon)^{p_0/2}$
either vanishes (if $p_0>0$) or blows up (if $p_0<0$). Yet the limit
$z=\epsilon\to z _0^-$ yielding $E_{\textnormal{\sc sing}}$ is
well-defined. Note also that the background integral in
eq.~(\ref{N-sig-K-tilde}) vanishes as $z=\epsilon\to z_0^-$, so the
reference background plays no part in the the computation of the
contribution $E_{\textnormal{\sc sing}}$ to the energy from the
singularity.  This was to be expected, since the cosmological constant
decouples near the singularity. We will come back to this below. The
total energy relative to the AdS background for the AdS-Kasner
solutions is the sum of eqs.~(\ref{E-inf}) and~(\ref{E-sin}),
\begin{equation}
  E_{\textnormal{\sc kas}} = E_\infty + E_{\textnormal{\sc sing}}
  = \frac{\ell^{n-1}V_{n-1}}{16\pi G z_0^n}\,(n-1)\,\,.
  \label{E-eon}
\end{equation}

\begin{figure}[ht]
\centering
  \includegraphics[width=0.55\textwidth]{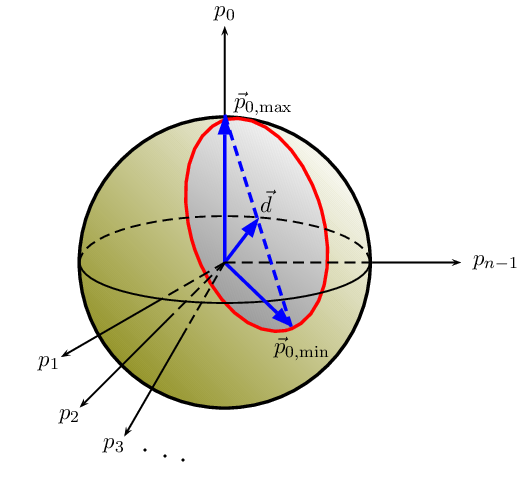}
  \caption{\sl Unit $(n\!-\!2)$--circle (in red) that results from the
    intersection of the unit $(n\!-\!1)$--sphere (in green)
    $\sum_ap^2=1$ with the plane $\sum_a p_a=1$ (in gray) in
    $\mathbf{R}^n$.}
  \label{fig-1}
\end{figure}

The asymptotic contribution $E_\infty$ to the energy, given by
eq. (\ref{E-inf}), grows linearly with $p_0$, and may be positive,
zero or negative. To determine its range, we must find the domain of
$p_0$.  Consider an arbitrary point
$\vec{p}=(p_0,p_1, \ldots,p_{n-1})$ in $\mathbf{R}^n$. In accordance
with eqs.~(\ref{conditions}), the values that the parameters $p_a$ in
the solutions~(\ref{metric-z}) may take lie on the $(n\!-\!2)$--circle
that results from the intersection of the plane $\sum_a\/p_a\!=1$ with
the unit $(n\!-\!1)$--sphere $\sum_ap_a^2\!=1$ (see
Figure~\ref{fig-1}). This circle is centered at
$\vec{d}=(\sfrac{1}{n},\, \sfrac{1}{n}\, \ldots, \sfrac{1}{n})$ and
has radius $\sqrt{1-(\sfrac{1}{n})}$.  The largest
possible value for $p_0$ is $1$, in which case all other $p_i$ vanish;
so the largest $p_0$ occurs at the point
$\vec{p}_{0,\rm max}=(1,0,\ldots,0)$ on the $(n\!-\!2)$-circle. The
smallest $p_0$ will occur at the point $\vec{p}_{0,\rm min}$ on the
circle located diametrically opposed to $\vec{p}_{0,\rm max}$, given
by $\vec{p}_{0,\rm min}\!= 2\vec{d}-\vec{p}_{0,\rm max}$. This implies
$\vec{p}_{0,\rm min}\!=\big(-1+(\sfrac{2}{n}),\,\sfrac{2}{n}\,\ldots,
\sfrac{2}{n})$, hence
\begin{equation}
-1+\frac{2}{n}\leq p_0\leq 1\,.
  \label{range-p0}
\end{equation}
It then follows that the asymptotic contribution $E_\infty$ in
eq.~(\ref{E-inf}) is negative for
$-1+(\sfrac{2}{n})\leq p_0 < \sfrac{1}{(2n-1)}$. The contribution
$E_{\textnormal{\sc sing}}$ to the energy from the siongularity, given
by eq.(\ref{E-sin}), is always positive and renders the total energy
$E_{\textnormal{\sc kas}}$ in eq.~(\ref{E-eon}) positive. Since $z_0$
is arbitrary for the AdS-Kasner solutions, if
$E_{\textnormal{\sc sing}}$ were not included in the energy but only
$E_\infty$ were considered, there would be gravitational fields with
arbitrarily negative energy.

The energies of the planar AdS black hole and the AdS soliton are
recovered from $E_\infty$ in eq.~(\ref{E-inf}) by setting $p_0\!=1$
and $p_0=0$,
\begin{align}
  E_{\textnormal{\sc bh}} &=\frac{ \ell^{n-1}V_{n-1}}{16\pi Gz_0^n}\,=(n-1)\, ,
  \label{E-bh}\\[4.5pt]
  \qquad
  E_{\textnormal{\sc sol}} & = -\frac{\ell^{n-1}V_{n-1}}{16\pi Gz_0^n}\,,
  \label{E-sol}
\end{align}
in agreement with the literature~\cite{HM-soliton}\,\cite{HWY}.  Note
that there are infinitely many solutions with $p_0\!=0$, hence with
the same negative asymptotic energy $E_\infty$ as the
soliton. Furthermore, there are solutions with nonzero $p_0$ whose
asymptotic energy $E_\infty$ is smaller that the soliton energy. The
point is that they all have a timelike singularity whose energy
$E_{\textnormal{\sc sing}}$ makes the total energy positive. From
eqs. (\ref{E-eon}) and (\ref{E-bh}) it follows that all the AdS-Kasner
solutions have the same total energy as the planar AdS black hole.

The reference background is introduced to have a well-defined finite
classical action. Once a reference background is chosen, we are not
free to remove it in regions of space at our will. Yet we have seen
that the reference AdS background does contribute to the energy of the
Kasner timelike singularity, which suggests that it is possible to
assign an energy to such singularities.  The metric near a Kasner
timelike singularity can be generically cast as
\begin{equation}
  ds^2_{n+1} (r\to 0) = dr^2  
  + b^2 \bigg[ - \Big(\frac{r}{r_0}\Big)^{2p_0} dt^2
  + \sum_{i=1}^{n-1} \Big(\frac{r}{r_0}\Big)^{2p_i} (dx^ i)^2\,\bigg]\, ,
    \label{Kasner-bis}
\end{equation}
with $r$ a radial coordinate centered at the singularity, $r_0$ a
characteristic length of the metric, and the $p_i$ satisfying the
Kasner conditions~(\ref{conditions}). The parameter $b$ can be hidden
in a redefinition of $t$ and $x^i$, but we keep it for later
convenience. Introduce a regulating timelike constant-$r$ surface
$\Sigma^r$, with $0<r\ll r_0$, around the singularity, with metric
\begin{equation}
  ds^2_{n} (\Sigma^r) = b^2 \bigg[ - \Big(\frac{r}{r_0}\Big)^{2p_0} dt^2
  + \sum_{i=1}^{n-1} \Big(\frac{r}{r_0}\Big)^{2p_i} (dx^ i)^2\,\bigg]\, .
    \label{Kasner-bis}
\end{equation}
Regarding $\Sigma^r$ as part of the spacetime boundary and
employing the same arguments as in ref.~\cite{Hawking-Horowitz}, the
contribution to the energy from the Kasner timelike singularity is
\begin{equation}
  E_{\textnormal{\sc ktl}} = -\frac{1}{8\pi G}\,\lim_{r\to 0}
  \int_{S^r_t} \!\!d^{n-1}y\, \sqrt{|\sigma|}\,N \kappa\,.
    \label{E-KTL}
\end{equation}
Here $S^r_t$ is the surface resulting from the intersection of
$\Sigma^r$ with a constant-$t$ slice $\Sigma_t$,
\hbox{$N=b\,(r/r_0)^{p_o}$} is the lapse, and $\kappa$ is the trace of
the extrinsic curvature of $S^r_t$ in $\Sigma_t$. Noting that
$\kappa=-(1-p_0)/r$, it is straightforward to arrive at
\begin{equation}
  E_{\textnormal{\sc ktl}} = \frac{b^n\,V_{n-1}}{8\pi G\,r_0}\,(1-p_0)\,,
  \label{E-KTL-res}
\end{equation}
which is positive and finite. For the Kasner metric metric in
eq.~(\ref{Kasner}) discussed earlier, $r_0=2\ell/n$ and $b=\ell/z_0$,
so $E_{\textnormal{\sc ktl}}$ reproduces to
$E_{\textnormal{\sc sing}}$ in eq.~(\ref{E-sin}). Of course, if the
singularity occurred in a different reference background, the
contribution of the background near the singularity would have to be
subtracted from the energy.

\section{Energy of the Schwarzschild metric with negative mass}

Consider the Schwarzschild geometry in $n+1$ dimensions,
\begin{equation}
  ds^2_{n+1} = -h(r)\,dt^2 + \frac{dr^2}{h(r)} + r^2d\Omega_{n-1}^2, \qquad
  h(r)=1 -\frac{m}{r^{n-2}} \,.
   \label{Sch-metric}
\end{equation}
As usual, $t$ and $r>0$ stand for the time and radial coordinates, and
$d\Omega_{n-1}^2$ is the metric on the unit $(n-1)$--sphere, whose
volume is $V_{n-1}={2\pi^{n/2}}/{\Gamma(n/2)}$. The reference
background is Minkowski space, to which spacetime approaches
asymptotically. Depending on the sign of $m$, two scenarios are
distinguished. If $m>0$, the metric describes a black hole of mass
\begin{equation}
  M_{\textnormal{\sc bh}} =\frac{(n-1) V_{n-1}}{16\pi G}\,m\,,
  \label{mass}
\end{equation}
with an event horizon at $r=m$.  Its energy is given by
$E_{\textnormal{\sc bh}}=E(r\to\infty)$, with~\cite{HM-soliton}
\begin{equation}
  E(r)=  -\frac{1}{8\pi G}\bigg(
  \int_{S^r_t} \!d^{n-1\!}x\, \sqrt{|\sigma|} N\kappa -
  \int_{\tilde{S}^r_t} \!d^{n-1\!}\tilde{x}\, \sqrt{|\tilde{\sigma}|}
                      {N} \tilde{\kappa}\bigg).
\label{E(r)}
\end{equation}
The surfaces $\tilde{S}^r_t$ and $S^r_t$ are now both the
$(n-1)$--sphere of radius~$r$, the lapse is $N=\sqrt{h}$, and the
traces $\kappa$ and $\tilde{\kappa}$ read
\begin{equation}
  \kappa=\frac{\sqrt{h}(n-1)}{r}\,,\quad
  \tilde{\kappa}=\frac{n-1}{r}\,.
  \label{misc}
\end{equation}
Substitution in eq.~(\ref{E(r)}) reproduces the well known result
$E_{\textnormal{\sc bh}}= M_{\textnormal{\sc bh}}$.

For $m<0$, there is no event horizon but a timelike singularity at
$r=0$, which is not of Kasner type. To regulate it, introduce a
constant-$r$ surface $\Sigma^r$, with $0<r=\epsilon\ll m$ the
regulating parameter. The energy is now the sum of the asymptotic
contribution $E_\infty=-|M_{\textnormal{\sc bh}}|$ and a contribution
from the singularity given by
$E_{\textnormal{\sc sing}}=E(\epsilon\to 0)$. To calculate the latter,
use $N=\sqrt{h}$ and that $\kappa$ and $\tilde{\kappa}$ are given by
minus the expressions in eqs.~(\ref{misc}), since the outward unit
normal vector to $S^\epsilon_t$ has opposite direction to that of
$S^\infty_t$. This gives\footnote{Note that $h(r)$ is positive for all
  $r$ if is $m,0$, so $\sqrt{h}$ is well-defined, and the limit
  $r=\epsilon\to 0$ is well-defined too.}
$E_{\textnormal{\sc sing}} = |M_{\textnormal{\sc bh}}|$, which is
positive. The total energy of the Schwarzschild black hole with
negative mass relative to Minkowski space is thus zero.

The $r=\epsilon\to 0$ limit of the second term in $E(r)$ in
eq.~(\ref{E(r)}) is zero, so again the reference background does not
modify the energy of the timelike singularity and we can associate to
the latter a positive energy
$E_{\textnormal{\sc sing}} = |M_{\textnormal{\sc bh}}|$ regardless of
the Minkowski background. Note however, that, while the energy of all
the AdS-Kasner spaces were positive, equal to the planar AdS black hole
energy, and larger than the energy of the AdS soliton, the energy of
the Schwarzschild black hole with negative mass is the same as that of
Minkowski space.

\section{Conclusion}

We have calculated the energy of timelike AdS-Kasner singularities and
of the timelike singularity at the center of a Schwarzschild black
hole with negative mass. These examples are by no means the most
general ones, yet they exhibit some interesting features. The energy
of the singularity depends on the spacetime geometry around it, is
positive and compensates for a negative energy from the asymptotic
region, rendering the total gravitational energy positive.  This
provides explicit realizations of the idea that geometries with a
timelike singularity are not good candidates for stable ground
states. This total energy, including the singularity's, can be thought
of as the internal energy of a solution, and it seems natural to use
it as the energy in thermodynamic considerations~\cite{Ong}.

One may think of viewing gravitational energy as a probe for
singularities in asymptotically AdS/flat spaces. We recall in this
regard that in recent years holographic complexity in the AdS/CFT
correspondence has been used as a probe for timelike
singularities~\cite{Barbon}\,\cite{Ren-2}. In particular, there is
certain parallelism between the treatment of the singularity here and
in the estimation of complexity as the action over a Wheeler-Dewitt
patch, since in the latter case calculating the contributions to the
action from the boundary and joints~\cite{LMPS} requires careful
regularization. The Authors of ref.~\cite{Ren-2} have computed
action-complexity in four dimensions for the AdS-Kasner metrics in
eq.~(\ref{metric-z}) using a regularization similar to that employed
here. The outcome of their calculation is that the singularity
augments complexity with respect to the AdS
background~\cite{Reynolds-Ross}.

\newlength{\bibitemsep}\setlength{\bibitemsep}{.3\baselineskip plus
  .05\baselineskip minus .05\baselineskip}
\newlength{\bibparskip}\setlength{\bibparskip}{0pt}
\let\oldthebibliography\thebibliography \renewcommand\thebibliography[1]{%
  \oldthebibliography{#1}%
  \setlength{\parskip}{\bibitemsep}%
  \setlength{\itemsep}{\bibparskip}%
}

\end{document}